\documentclass[usenatbib,usegraphicx,useAMS]{mn2e}
\usepackage{amsmath}
\usepackage{amssymb}

\title[Triple Disruptions in the Galactic Centre]{Triple Disruptions in The Galactic Centre: Captured and Ejected Binaries, Rejuvenated Stars, and Correlated Orbits}

\author[Ginsburg \& Perets]{Idan Ginsburg\thanks{E-mail:
idan.ginsburg@dartmouth.edu} \& Hagai B.
Perets\thanks{E-mail:hperets@cfa.harvard.edu}\\Department of Physics and 
Astronomy, 6127 Wilder Laboratory, Hanover, NH 03755, USA
\\Harvard-Smithsonian Center for
Astrophysics, 60 Garden St., MS 51, Cambridge, MA 02138, USA\\}

\begin{document}
\maketitle

\begin{abstract}
The disruption of a binary star by a massive black hole (MBH) typically leads to the capture of one component around the MBH and the ejection of its companion at a high velocity, possibly producing a hypervelocity star. The high fraction of observed triples ($\sim10$ \% for F/G/K stars and $\sim50$ \% for OB stars) give rise to the possibility of the disruption of triples by a MBH. Here we study this scenario, and use direct $N$-body integrations to follow the orbits of thousands of triples, during and following their disruption by a MBH (of $4\times10^6$ M$_\odot$, similar to the MBH existing in the Galactic Centre; SgrA$^*$). We find that triple disruption can lead to several outcomes and we discuss their relative frequency. Beside the ejection/capture of single stars, similar to the binary disruption case, the outcomes of triple disruption include the ejection of hypervelocity binaries; capture of binaries around the MBH; collisions between two or all of the triple components (with low enough velocities that could lead to their merger); and the capture of two or even three stars at close orbits around the MBH. The orbits of single stars captured in a single disruption event are found to be correlated. The eccentricity of the mutual orbits of captured/ejected binaries is typically excited to higher values. Stellar evolution of captured/ejected binaries may later result in their coalescence/strong interaction and the formation of hypervelocity blue stragglers or merger remnants in orbits around SgrA*. Finally, the capture of binaries close to the MBH can replenish and increase the binary frequency near the MBH, which is otherwise very low.
\end{abstract}


\section{Introduction} \label{INT}
First theorized by \citet{Hills:88}, a close interaction by a binary star system and massive black hole (MBH) can produce a hypervelocity star (HVS) that has sufficient velocity to escape the gravitational pull of the galaxy where it originated, and leave its companion orbiting near the MBH (\citealt{Gould-Quillen:03}). Since the discovery of the first HVS in 2005 \citep{Brown:05}, at least 16 HVSs have been observed in the Milky Way (\citealt{Edelmann:05}; \citealt{Hirsch:05}; \citealt{Brown:06a}; \citealt{Brown:06b}; \citealt{Brown:07b}; \citealt{Brown:09b}). There are a number of mechanisms that can produce HVSs. The Hills' mechanism has been well studied and appears to be the most likely source of HVSs in the Milky Way (\citealt{Yu-Tremaine:03}; \citealt{Ginsburg:1}; \citealt{Perets:07}; \citealt{Perets:09a}; \citealt{Madigan:09,Madigan:10}). Other scenarios include the interaction of stars with stellar black holes \citep{Oleary-Loeb:08}, and the inspiral of an intermediate mass black hole (IMBH) (\citealt{Hansen:03}; \citealt{Yu-Tremaine:03}; \citealt{Levin:06}; \citealt{Sesana:09}). It was also suggested that a binary star system could be ejected as a hypervelocity binary through interactions with an IMBH (\citealt{Lu:07}; \citealt{Sesana:09}). \citet{Perets:09b} (hereafter Perets I) expanded upon the binary hypervelocity scenario and showed that massive hypervelocity binaries could be ejected through a triple disruption by a MBH.

Some HVSs have been found at such distances from their likely origin at the Galactic Centre (GC) (see \citealt{Brown:09b}), that despite their high velocities, the main sequence (MS) lifetime of these stars is less than their flight time. Perets I suggested that this discrepancy can be solved if the stars were originally ejected as binaries, and later coalesced to form single rejuvenated stars (blue stragglers).

Stellar dynamics (mostly of young B-stars observed close to SgrA$^*$) within the central arcsecond of the GC provide overwhelming evidence for the existence of a central MBH with mass $\sim 4\times10^6M_{\odot}$ (e.g. \citealt{Scho:03}; \citealt{Reid-Brunthaler:04}; \citealt{Ghez:05}; \citealt{Ghez:08}; \citealt{Gillessen:2009}). Aside from producing HVSs via the Hills' mechanism, the tidal breakup of a binary could lead to the capture of a star in a close orbit around the MBH \citep{Gould-Quillen:03}, possibly explaining the origin of young B-stars around SgrA$^*$ (\citealt{Perets:07,Madigan:09,Perets:09a,Perets+:09a,Madigan:10,Perets:10}). \citet{Ginsburg:2} showed that the tidal breakup of a stellar binary by the MBH can also lead to a collision that ends with coalescence. Such mergers may be a source of young stars for the Galactic Centre (\citealt{Antonini:10a}; \citealt{Antonini:10b}). 

The tidal binary break-up scenario provides a wide variety of outcomes, which were studied at length. Here we expand the study to the disruption of triple stars, which given the observed high triple frequency ($\sim10$ \% for F/G/K stars and $\sim50$ \% for OB stars; \citealt{Raghavan:2010,Evans:2011}), are highly likely to occur. This possibility was discussed in general terms analytically (\citealt{Lu:07}; Paper I). We examine the results of over 2000 $N$-body simulations of triple disruptions by a MBH, enabling us to study their various outcomes quantitatively, and compare them with binary disruptions. We discuss these outcomes and their relative frequency, as well as their long term evolution. Note however, that given the large phase space of possible initial conditions, our simulations are not exhaustive, and cover only a limited range of conditions.

In section 2 we describe the codes and simulation parameters we used. In section 3 we discuss the triple disruptions scenario. In section 4 we discuss the formation of a blue straggler (BS) and consequently the formation of a hypervelocity binary (hereafter HVB) from a triple disruption.

\section{Computational Method} \label{CM} 
In our study we have used the N-body code written by \citet{Aarseth:99}, which is freely available for download \footnote{http://www.ast.cam.ac.uk/$\sim$sverre/web/pages/nbody.htm}.  We adopted a small value of $10^{-8}$ for the accuracy
parameter $\eta$, which determines the integration step through the relation $dt=\sqrt{{\eta F}/(d^2F/dt^2)}$ where $dt$ is the timestep and $F$ is the force.  The softening parameter, \emph{eps2}, which is used to create the softened point-mass potential, was set to zero. We treat the stars as point particles and ignore tidal and general relativistic effects
on their orbits, since these effects are small at the distance ($\sim10$AU; over the short dynamical timescale of the encounter) where the triple or binary system is tidally disrupted by the MBH.

We set the mass of the MBH to $M=4\times 10^6M_{\odot}$. For our initial run, we set the mass of each star to $m=4M_{\odot}$ within the triple system; comparable to the currently observable hypervelocity stars.  For the next set of simulations, we used the same parameters as that of the triple systems, however we used a binary system with mass $m_1 = 8M_{\odot}$ and $m_2 = 4M_{\odot}$, equivalent to the combined mass of the triple system. Our initial simulations were planar, however we also ran over 500 simulations with an inclination angle of 60 degrees. Most of our simulations assumed initial orbital eccentricities of $e=0$. For comparison, we ran a set of over 500 simulations with $e=0.4$. All runs start with the centre of the circular triple system located 2000 AU ($=10^{-2}$pc) away from the MBH along the positive y-axis. This radius is larger than the binary size or the distance of closest approach necessary to obtain the relevant ejection velocity of HVSs, making the simulated orbits nearly parabolic.  We used the same initial distance for all runs to make the comparison among them easier to interpret as we varied the distance of closest approach to the MBH or the relative positions of the stars within the system.

\begin{figure}
\begin{center}
\includegraphics[width=3in]{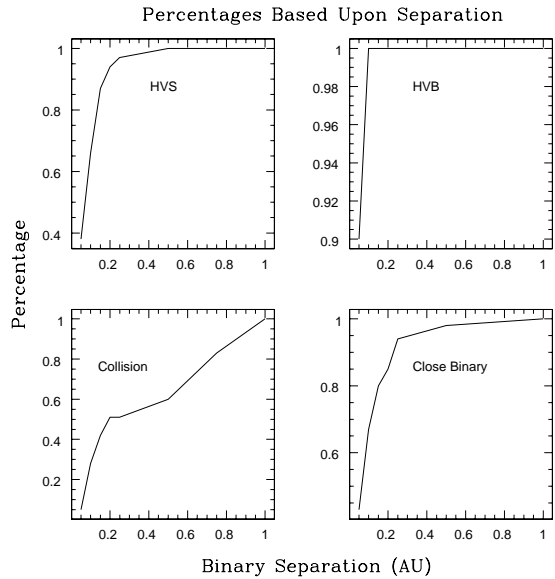}
\end{center}
\caption{This cumulative plot shows the fraction of HVSs, HVBs, collisions, and tight binaries produced from the disruption of a triple system by the MBH given an initial binary separation (AU). These results are based upon our initial simulations with each star of mass $4M_{\odot}$. The production of HVSs, HVBs, and close binaries all drop exponentially with increasing binary separation. In particular, the production of HVBs occurs only when the initial binary separation is very tight. The rate of collisions stays fairly constant with increasing binary separation. Note that for these simulations the initial separation of the third star is 5 times the binary separation.}
\label{lsq}
\end{figure}

For most of our simulations involving triple systems, we chose the inner binary to have semimajor axis of length $a_o=0.05, 0.1, 0.15$, and $0.2$ AU, and left the second semimajor axis at $5a_o$ in order to keep the binary as tight as possible. However, we did run simulations with larger $a_o$ as well as simulations with semimajor axis length greater than $5a_o$ to see how this increase affects the disrupted systems (see Figure \ref{lsq} and Table \ref{tab_semimajor}). Such a distribution of semimajor axis lengths allows us to test the range of hypevelocity stars and hypervelocity binary production. Wider binaries give lower ejection velocities \citep{Hills:1991,Bromley:06}.  Much tighter binaries would not be easily disrupted by the black hole, or may coalesce to form a single star before interacting with the MBH. The size of a MS star of a few solar masses is $\sim 0.01$AU, and so binaries tighter than $\sim 0.02$AU are precluded because the two stars will shortly merge. 

In the Galactic disk, $\sim0.75$ of all stellar systems with O/B and $\sim0.3$ of systems with F/G/K primaries, are binaries or higher multiplicity systems (see e.g. \citealt{Raghavan:2010,Kobulnicky:2007}), with roughly equal probability 
per logarithmic interval of separations, $dP/d\ln(a)=const$ (e.g. \citealt{Abt:83}; \citealt{Heacox:98}; \citealt{Larson:03}). F/G/K binaries have a log-normal distribution of periods \citep{Raghavan:2010}. For such a distribution, $\sim0.3-0.4$ of all OB binaries have  $< 1$ AU separations (see discussion of binary and triple fractions in Paper I), where F/G/K binaries are likely to only have $\sim0.05$ of the systems in this range.

The initial phase of the stellar orbit plays a crucial role in the outcome \citep{Ginsburg:1}. Therefore, we sampled cases with initial phase values of $0-180^o$ in increments of $15^o$.  As initial conditions, we gave the binary system no radial velocity but a tangential velocity with an amplitude such that our effective impact parameter is $\sim 5-10$ AU. We expect no HVSs to be produced at substantially larger impact parameters \citep{Ginsburg:1}.

We also examine cases for triple systems with tight inner binaries ($0.05$ and $0.1$ AU) and masses in the range $1,2,3,5$ and $6 M_{\odot}$ per star. Note that a triple system of $M_{triple} \leq 3M_{\odot}$ with a close outer binary is far less likely than one where the primary is of higher mass $m \sim 5M_{\odot}$  (\citealt{Tokovinin:06}; \citealt{Tokovinin:8}). 
Upon completion of all our simulations, we evolved the HVBs using the binary-star evolution (BSE) code written by \citet{Hurley:02}, and compared the results with the evolution of a single star of similar mass using the single-star evolution (SSE)  code \citep{Hurley:00}. Our results show that a HVB can coalesce well after its ejection, and be observable as a MS star even after $4-5$ times the lifetime of a single star of similar total mass ($m \sim 7M_{\odot}$). Such a HVB could therefore be observed at far larger distances than accessible to a single star with the same total mass and velocity, which would evolve off the MS.

\section{The Triple Disruption Scenario} \label{TDS}

\begin{figure*}
\begin{center}
\begin{tabular}{ccc}
\includegraphics[width=0.32\textwidth]{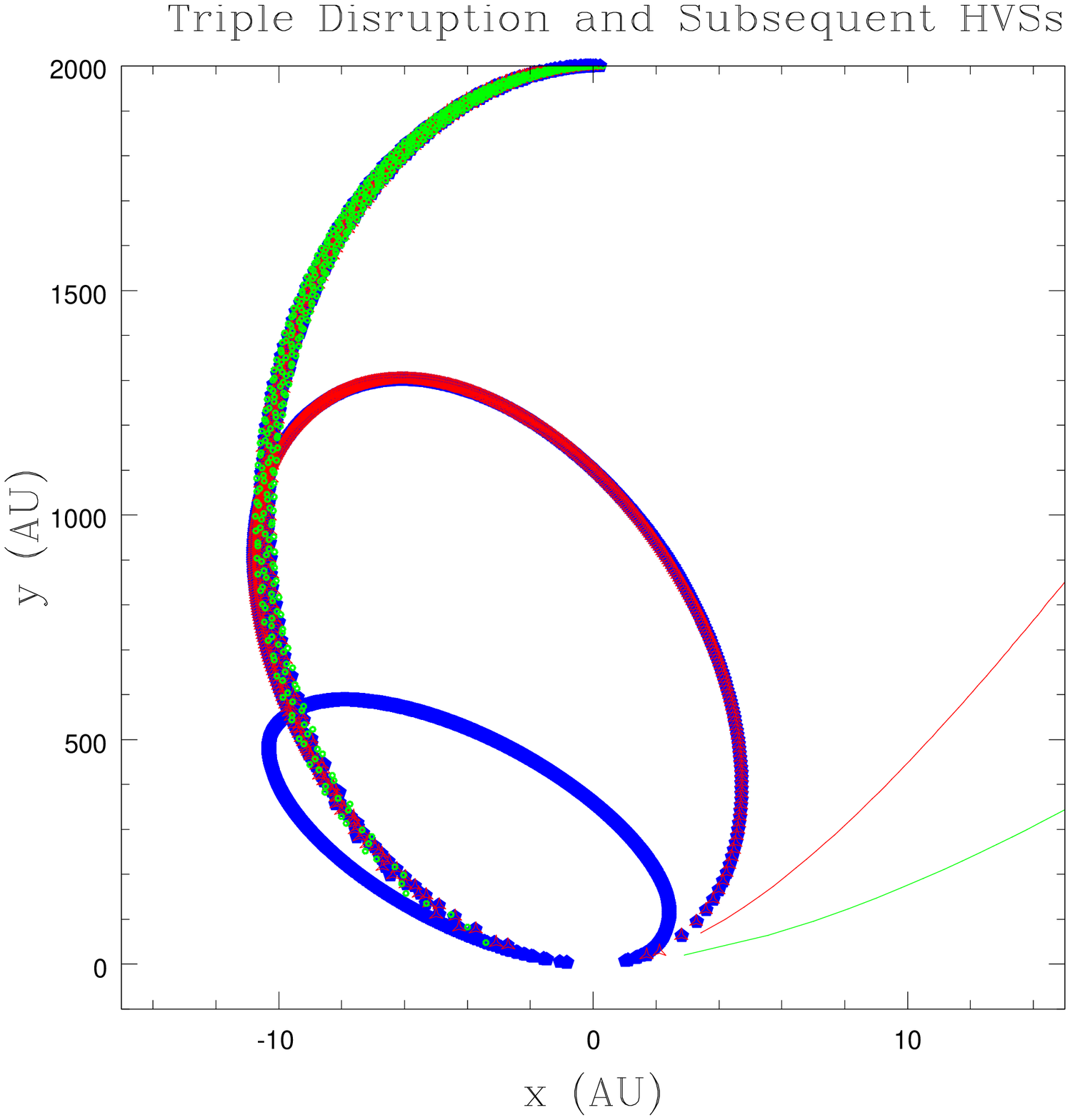}
&
\includegraphics[width=0.32\textwidth]{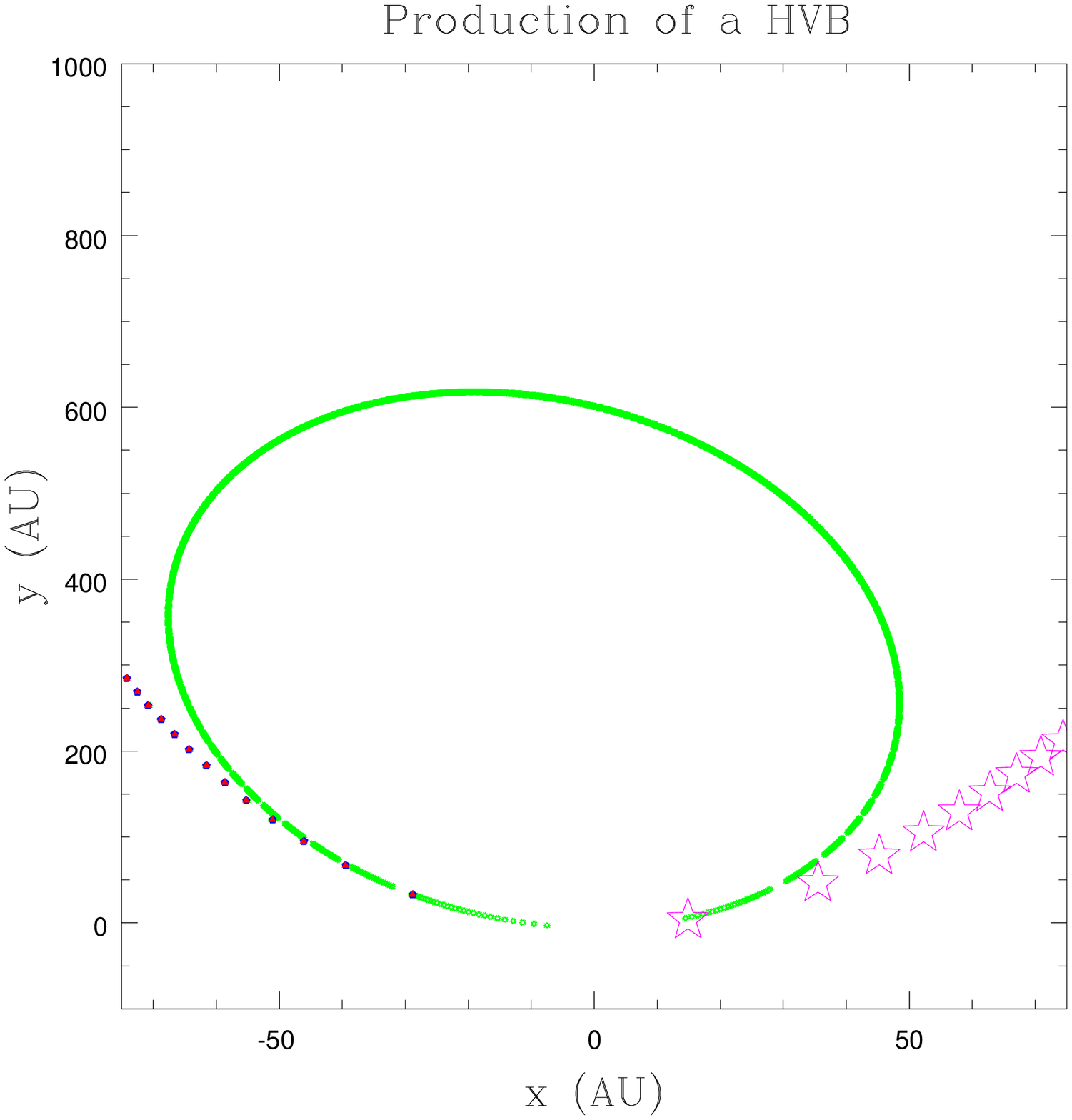}
&
\includegraphics[width=0.32\textwidth]{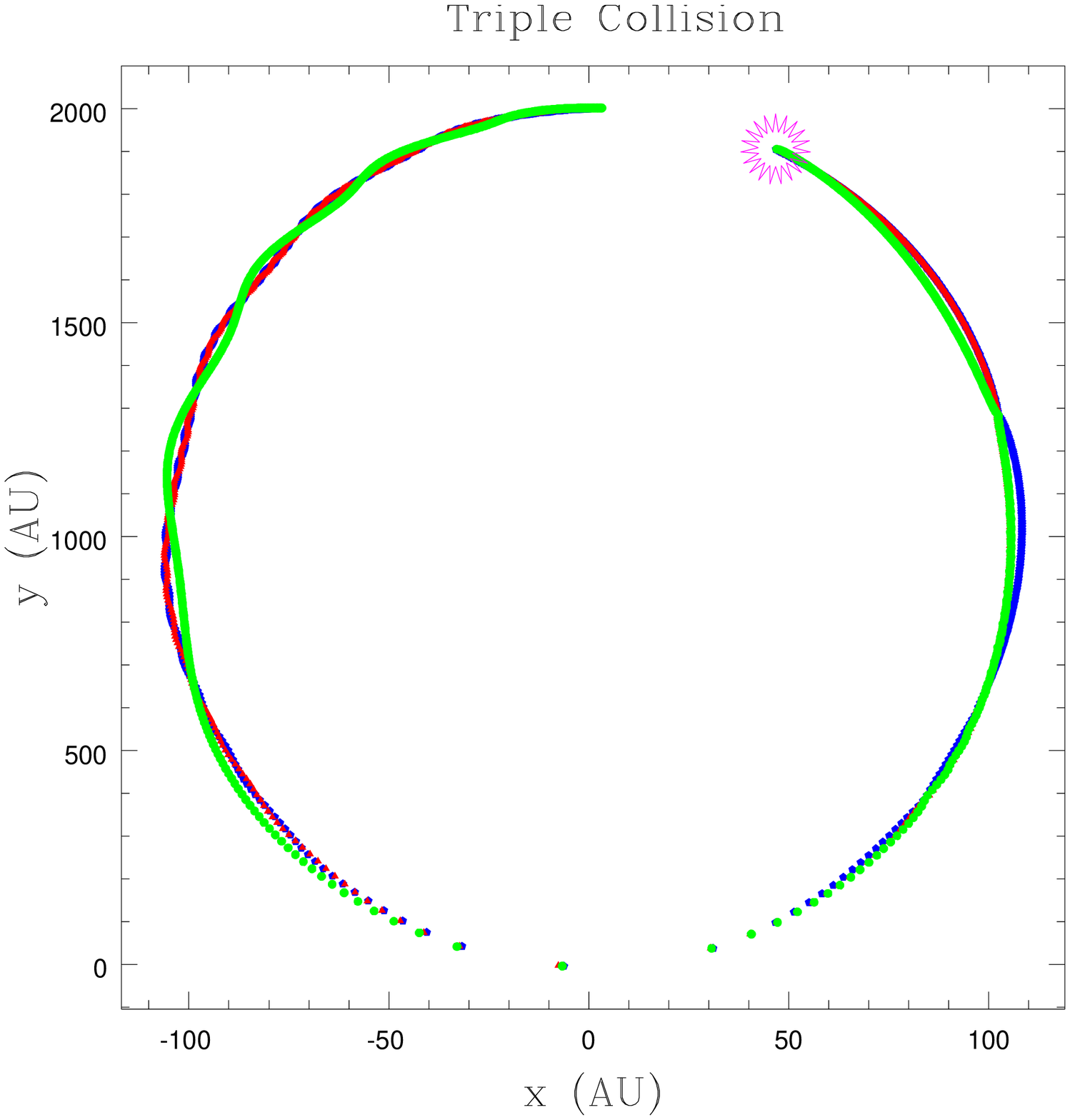}
\\
\end{tabular}
\end{center}
\caption{The three images show the evolution of a triple system after being disrupted by the MBH. In each scenario the system is initially located $2000$ AU along the positive y-axis. The separation between the inner binary is $a=0.1$ AU, and between the inner binary and third star is $0.5$ AU. The stars of the inner binary are denoted by the red and blue colors respectively, and the outer binary by the color green. Each star has a mass of $4M_{\odot}$. In the first case (left image) the triple system produced a HVS upon being disrupted and the ejected star was the outer binary (the HVS's path is denoted by the solid green line). The remaining binary orbited the MBH until it was also disrupted and a second HVS produced. The path of this second HVS is denoted by the solid red line. The remaining star (blue) is in a highly eccentric orbit around the MBH. The middle image shows the production of a hypervelocity binary (HVB). In this case, the outer star remains in a highly eccentric orbit around the MBH, while the inner binary (denoted by the magenta star) is ejected. In the right image the triple system is perturbed by the MBH, and before completing an orbit all three stars collide with $v<v_{esc}$ where the escape speed for a star of a few solar mass is $\sim 500 kms^{-1}$. It is likely that in such a scenario the three stars will coalesce.}
\label{outcome}
\end{figure*}

Given a binary system with stars of equal mass $m$ separated by a distance $a$ and a MBH of mass $M\gg m$ at a distance $b$ from the binary, tidal disruption would occur if $b\la b_{\rm t}$ where

\begin{equation}
\frac{m}{a^3} \sim \frac{M}{{b^3_{\rm t}}}.
\end{equation}

\noindent The distance of closest approach in the initial plunge of the binary towards the MBH can be obtained by angular momentum conservation from its initial transverse speed $v_{\perp}$ at its initial distance from the MBH, $d$,

\begin{equation}
v_{\perp}d = \left(\frac{GM}{b}\right)^{1/2}b .
\end{equation}

\noindent The binary will be tidally disrupted if its initial transverse speed is lower than some critical value,

\begin{equation}
v_\perp\la v_{\perp,\rm crit} \equiv {(GMa)^{1/2}\over d}\left({M\over m}\right)^{1/6}= 10^2 {a_{-1}^{1/2} \over m_{0.5}^{1/6} d_{3.3}} ~{\rm {km~s^{-1}}},
\label{eq:crit}
\end{equation}

\noindent where $a_{-1}\equiv ({a}/{0.1~{\rm AU}})$, $d_{3.3}=(d/2000~{\rm AU})$, $m_{0.5} \equiv (m/3M_{\odot})$, and we have adopted $M=4\times 10^6M_{\odot}$.  For $v_\perp\la v_{\perp,\rm crit}$, one of the stars receives sufficient kinetic energy to become unbound, while the second star is kicked into a tighter orbit around the MBH.  The ejection speed, $v_{\rm ej}$, of the unbound star can be obtained by considering the change in its kinetic energy $\sim v\delta v$ as it acquires a velocity shift of order the binary orbital speed $\delta v \sim \sqrt{Gm/a}$ during the disruption process of the binary at a distance $\sim b_t$ from the MBH when the binary centre-of-mass speed is $v\sim \sqrt{GM/b_t}$ \citep{Hills:88,Yu-Tremaine:03}. At later times, the binary stars separate and move independently relative to the MBH, each with its own orbital energy.  For $v\la v_{\perp,\rm crit}$, we therefore expect


\begin{align}
v_{\rm ej} \sim \left[\left({\frac{Gm}{a}}\right)^{1/2}\left({\frac{GM}{b_{\rm t}}}\right)^{1/2}\right]^{1/2} \nonumber\\ = 1.7 \times 10^3 m^{1/3}_{0.5}a^{-1/2}_{-1} ~{\rm km~s^{-1}}.
\label{eq:model}
\end{align}

For a triple system, the possible outcomes are (i) the stars are disrupted and separately orbit the MBH; (ii) a HVS is produced and the remaining binary orbits the MBH (the binary might then be disrupted in such a way that a second HVS is produced); (iii) a HVB is produced and the remaining star orbits the MBH; (iv) a collision occurs between either two or all three stars; (v) the stars are disrupted but a close binary remains around the MBH. During a collision, if the impact velocity is less than $v_{esc}$, the stars will likely coalesce. If a close binary has semimajor axis length such that mass transfer can occur, then a possible outcome is a BS. Scenarios ii-iv are illustrated in Figure \ref{outcome}. Table \ref{tab_compare} summarizes the statistical results from our initial runs using stars of $4M_{\odot}$ with $a \leq 0.2$ AU.

\begin{table}
\begin{center}
\begin{tabular}{|r|r|r|r|r|r|}
\hline
Outcome For 3 Stars &P($a\leq 0.2$ AU)\\
\hline
HVS + 2 Capture& 0.26${\pm}$0.03& \\
HVS + Binary Capture& 0.02${\pm}$0.01& \\
2 HVS + 1 Capture& 0.04${\pm}$0.01& \\
HVB& 0.05${\pm}$0.01& \\
Collision& 0.04${\pm}$0.01& \\
Binary Capture (no HVS) & 0.11${\pm}$0.02& \\
3 Capture& 0.48${\pm}$0.05& \\
\hline
Outcome For 2 Stars &P($a\leq 0.2$ AU)\\
\hline
$4M_{\odot}$ HVS& 0.20${\pm}$0.03& \\
$8M_{\odot}$ HVS& 0.13${\pm}$0.02& \\
Collision& 0.01${\pm}$0.01& \\
2 Capture&0.54${\pm}$0.05& \\
Binary Capture&0.12${\pm}$0.02& \\
\hline
\end{tabular}
\end{center}
\caption{The first section of the table shows the probability of various outcomes for a triple system composed of $4M_{\odot}$ stars upon disruption by the MBH. Here the binaries are all of semimajor axis $a \leq 0.2$ AU. The third star is at a distance of $5a$ for all cases. Each row shows the values obtained from our initial simulations with their corresponding Poisson errors. The first row shows the probability that a single HVS is produced, and the remaining stars orbit the MBH independently. The second row shows the probability that a HVS is produced, and the remaining stars orbit the MBH as a binary. The third row shows the probability that 2 HVSs are produced, and consequently the remaining star orbits the MBH. The fourth row shows the probability of producing a HVB. The fifth row shows the probability of a collision for either 2 or 3 stars (note that a collision here does not necessarily imply coalescence). The sixth row shows the probability for a binary to be captured. In this case there is no HVS produced, and the remaining star orbits the MBH in an independent orbit. The seventh row shows the probability of the MBH capturing all 3 stars. In this scenario the final orbit of each star is independent of the others. The second section of the table shows the probability of various outcomes for a binary system composed of $4M_{\odot}$ and $8M_{\odot}$ stars. The binary separation is similar to the binary separation for the triple scenario, where $a \leq 0.2$ AU. The first row shows the probability of producing a $4M_{\odot}$ HVS, while the second row shows the probability of producing a $8M_{\odot}$ HVS. The third row shows the probability of a collision. The fourth row shows the probability that both stars are captured by the MBH in independent orbits. The fifth row shows the probability that both stars are captured by the MBH as a binary.}
\label{tab_compare}
\end{table}

\subsection{Comparison With A Binary System}

In order to verify whether the initial disruption of a triple system behaves differently than a binary disruption, we ran simulations of binary disruptions by the MBH with the same parameters as those in our triple systems. The innermost star was given a mass of $8M_{\odot}$, the companion $4M_{\odot}$, and all other properties (semimajor axis, inclination, impact parameter) remained the same. Although a binary system can not produce a HVB, most of the other outcomes are statistically similar to those of the triple systems. Our results are summarized in Table \ref{tab_compare}. We conclude that the initial disruption of a tight triple system or binary system behave similarly.

\subsection{Inclination and Eccentricity}

Most of our simulations were run with planar configurations (relative inclination of the binary and its trajectory around the MBH of $i=0^\circ$), and with initial binary orbital eccentricity of $e=0$. In order to explore the effects of the inner binary properties on the outcomes, we have run additional simulations with mutual binary inclination of $i=60^\circ$ (on circular orbits; $e=0$) as well as simulations of eccentric binaries, $e=0.4$ (with planar orbits; $i=0^\circ$) while keeping other parameters unvaried. The orbital inclination of the captured stars is determined primarily from their initial inclination (see \citealt{Zhang:10} for a discussion on binary HVSs, the case is similar for triple disruptions). Our results are summarized in Table \ref{tab_ecc}. While roughly similar to our results from Table \ref{tab_compare}, we find that for a high inclination and higher eccentricity, more HVSs are produced. In addition, higher inclination leads to more collisions.

\begin{table}
\begin{tabular}{|r|r|r|r|r|r|}
\hline
outcome &P($i = 60^o$) & P($e=0.4$) \\
\hline
HVS + 2 Capture& 0.49${\pm}$0.05& 0.38${\pm}$0.04\\
HVS + Binary Capture& 0${\pm}$0& 0${\pm}$0\\
2 HVS + 1 Capture& 0.08${\pm}$0.02& 0.02${\pm}$0.01\\
HVB& 0.02${\pm}$0.01& 0.05${\pm}$0.01\\
Collision& 0.07${\pm}$0.02& 0.02${\pm}$0.01\\
Binary Capture (no HVS) & 0.04${\pm}$0.01& 0.10${\pm}$0.02\\
3 Capture& 0.3${\pm}$0.04& 0.43${\pm}$0.05\\
\hline
\end{tabular}
\caption{Probability of various outcomes for a triple system composed of $4M_{\odot}$ with orbital inclination 60 degrees (left column), and for a system with initial eccentricity $e=0.4$ (right column) upon disruption by the MBH. For both scenarios the semimajor axis of the third star was $5a$ where $a$ is the semimajor axis of the inner binary with $a \leq 0.2 AU$ for all simulations. Statistically, the results are comparable to Table \ref{tab_compare}, with some notable differences. In both cases we have more HVSs produced, and less complete captures of all stars. Also for the first case we have more collisions and less binary capture rates which are due to Kozai-like effects. Both cases are meant to illustrate possible situations, and are not conclusive in and of themselves.}
\label{tab_ecc}
\end{table}

\section{The outcomes of triple disruptions}
\label{outcomes}

In the following, we discuss the possible outcomes of triple disruptions and their potential observable products.

\subsection{HVB}

The Hills' mechanism which is responsible for the production of HVSs can also produce a HVB through the disruption of a triple system. In order to produce a HVB the triple system will need to have $M_{triple} \sim 6-12M_{\odot}$ and $a_{bin} < 1$ AU (Perets I). Although the fraction of such binaries is not well known, it is estimated to be $\sim 0.05$ (\citealt{Fekel:81}; Perets I). A HVB produced this way will be a close or contact binary. Consequently, it is possible that most of the observed contact binaries were produced through evolution in triple star systems (\citealt{Pribulla-Rucinski}; \citealt{Fabrycky-Tremaine}). Our simulations confirm that all HVBs are close or contact binaries with orbital period of $\sim 2$ days. Such close binaries are necessary for the formation of BSs \citep{Lu:10}. Furthermore, there is evidence that suggests nearly all observed BSs are formed in binary systems \citep{Leigh:07}.

In our simulations we primarily concentrated on tight triple systems ($a_o\leq 0.2$ AU) with stars of mass $4M_{\odot}$, however we also ran simulations to see what affect mass and semimajor axis length has on the HVB and HVS production rates. We found that an inner semimajor axis length of $a=0.2$ AU or less is necessary for the production of HVBs, so long as the outer semimajor  axis is $5a$ (see Table \ref{tab_semimajor}). Similarly, the rate of HVS production drops dramatically once the semimajor axis of the inner binary is $> 0.2$ AU. Keeping the semimajor axes ratio at 5, and the inner binary tight, we found that increasing the mass of the system also increases the likelihood of a HVB.

For some bound HVSs (see \citealt{Brown:07b}), and also HVS HE 0437-5439 \citep{Edelmann:05} there exists a discrepancy between their kinematics and apparent ages. A number of theories have been proposed to explain the fact that their lifetime is longer than expected. Our simulations show that although rare (see Table \ref{tab_compare}), HVBs can be produced via the disruption of triple systems with tight binaries. Taking the parameters (orbital period and eccentricity) from our $N$-body simulations and evolving the HVB using BSE, we found that the two $4M_{\odot}$ stars will form a common envelope, and via mass transfer one star will reach a mass of $\sim 7M_{\odot}$. The second star typically goes supernova (massless), and the remaining star on average stays on the MS for $\sim 200$ Myr. Typically, a star of $7M_{\odot}$ will live for no longer than $\sim 50$ Myr, therefore the HVB evolves into a runaway BS. Our BSE simulations show that short periods ($\leq 10$ days) but high eccentricity ($e \geq 0.6$) will produce a star of $\sim 7M_{\odot}$ with lifetime $\leq 50$ Myr. The average orbital period of our HVBs is $\sim$ 2 days, and the eccentricity is typically low, with an average value of $\sim 0.3$. Such scenarios allow for rejuvenation and can explain the unexpected longevity of such HVSs.

\subsection{Close Binary Around SgrA*}

The disruption of a triple system is three times more likely to produce a captured binary than a HVB (see Table \ref{tab_compare}). We define a captured binary as a binary with semimajor axis $\leq$ 2 AU orbiting the MBH.  Our simulations show that if two stars remain bound together after their initial tidal interaction with the MBH, then they almost always have semimajor axis  $\leq$ 2. Otherwise the system is disrupted and the stars orbit the MBH independently and mass transfer is not possible. The orbital parameters of such a binary are similar to those of a HVB, and consequently the evolution will in most cases be identical to that of a HVB. Our average period for a close binary was $\sim$ 10 days, and the eccentricity $\sim$ 0.4. Such values lead to an average lifespan of $\sim$ 190 Myr for a BS of mass $\sim 7M_{\odot}$. As in the case for a HVB, the eccentricity greatly affects the lifetime of the BS. Tidal stripping may also affect the evolution of the binary, and it has been suggested that the S-stars orbiting SgrA* are tidally stripped AGB stars \citep{Davies-King}, however there is no confirmation of any stripped stars at the Galactic Centre, and current evidence indicates that S2 is a genuine early B star of mass $14-20M_{\odot}$ \citep{Martins:08}. Although not all of the stars orbiting SgrA* could come from triple disruptions, it is possible that some of the stars are blue stragglers that evolved from triple disruptions.

\subsection{Collisions}

Direct collisions are one of the primary mechanism for BS formation, and in dense environments collisions often involve more than two stars \citep{Fregeau:04}. Under some circumstances a triple system is disrupted in such a way that either two or all three stars collide. Assuming that a binary is disrupted at a random angle and ignoring gravitational focusing, the probability for a collision is 

\begin{equation}
P \sim \frac{4R_*}{2\pi R_{sep}}
\end{equation}

\noindent where $R_{\star}$ is the radius of the star (assuming radii are approximately equal) and $R_{sep}$ is the average separation of the stars. The stars will merge if the relative velocity of impact is less than the escape velocity from the surface of the star ($\sim 500 kms^{-1}$). In our simulations around 15 per cent of all collisions have impact velocities low enough to allow two stars to coalesce, and roughly 4 per cent allow all three stars to coalesce. A simple estimation for the impact velocity comes from conservation of energy,

\begin{equation}
E = \frac{1}{2}\frac{m_1m_2}{m_1+m_2}\dot r^2 - \frac{Gm_1m_2}{r} = constant
\end{equation}

\noindent which yields the relative impact velocity for two stars

\begin{equation}
v_f = [2G(m_1+m_2)(\frac{1}{a_{min}} - \frac{1}{a})]^{1/2}.
\end{equation}

\noindent The equations are nearly the same for three stars, with an added $m_3$. Equation (7) does not take tidal forces into account nor gravitational focusing. The rate of HVS production in the Milky Way is $\sim 10^{-5} yr^{-1}$ \citep{Brown:06b}, and the total collisional rate is $\sim 10$ per cent \citep{Ginsburg:2}. Given that mergers occur in $\sim 20$ per cent of all collisions, the expected production rate of such S-stars is once very 5 Myr. The lifetime of a massive star of mass $\sim 14M_{\odot}$ is roughly 10 Myr. Therefore mergers can not account for a majority of the S-stars, but may account for some and can help explain the longer than expected lifetimes of these stars.

\section{Evolution of captured and ejected binaries} \label{FBS}

Following the triple disruption, captured binaries could evolve through short (orbital) and long (secular) term dynamical evolution induced by the MBH on other stars in their environment. They may also evolve later through binary stellar evolution. HVBs may similarly evolve through binary stellar evolution, but their mutual orbits are unlikely to be affected by interactions with other stars which are scarce once a HVB is ejected from the GC.

\subsection{Dynamical evolution of captured binaries}

The dynamical channels could lead to the disruption of the binaries or to their merger/collision. In many cases a triple disruption is followed by the capture of a binary, which itself is disrupted shortly after to leave two bound components around the MBH which are unbound to each other (see Table \ref{tab_compare} and Table \ref{tab_ecc}). This possibility leads to both binary components orbiting the MBH in individual, but correlated orbits. Over longer timescales, relaxation processes such as resonant relaxation could change the orbits of the captured stars such that their inclination and eccentricity no longer seem correlated \citep{Perets+:09b}. 

In other cases (see \S\ref{outcomes}), following the triple disruption, a binary is captured but collides to form a merger product. When binaries are captured with high inclination between their mutual orbits and the orbit around the MBH, they are more likely to collide, as secular Kozai-like effects can excite their eccentricity inducing the merger (Paper I; \citealt{Antonini:10a}; \citealt{Antonini:10b}). Indeed, we find that triple disruptions of inclined triples (i.e. leading to the capture of inclined binaries) show a higher rate of collisions at the expense of the binary capture rate, which is decreased (see Table \ref{tab_ecc}). The collision/merger products could potentially be observed as atypical, more massive S-stars, possibly with high rotation velocities.

\subsection{Stellar evolution of captured and hypervelocity binaries}  

Captured binaries which did not collide shortly after their capture, as well as ejected HVBs, may later evolve through stellar evolution. Such evolution may lead to their merger through mass transfer and common envelope evolution, or to the production of compact binaries in the GC; or in the case of HVBs, to hypervelocity merger products and compact binaries.

First discovered by \citet{Sandage}, a BS is a MS star with an apparently abnormally long lifetime. BSs are massive stars that lie above the turnoff region in the blue end of the color-magnitude diagram. If BSs were normal MS stars, they would have evolved away from the MS.  There are a number of theories proposed to explain the unexpected longevity of BSs. One theory involves stellar collisions and coalescence, another mass transfer from a binary companion, and a third is from induced merger/collision through evolution in triple systems (see \citealt{Stryker} for a review of the two former theories and \citealt{Perets+:09b}; Paper I; and \citealt{Antonini+Perets}, regarding the evolution in triples). 

Using the BSE binary stellar evolution code by Hurley \citep{Hurley:02}, we evolved two stars of $m_1 = m_2 = 4M_{\odot}$ with period $P \sim 10$ days. Such a system evolves into a $7M_{\odot}$ star with a companion of $m < 1M_{\odot}$.  The lifetime of the evolved $7M_{\odot}$ star is $\sim$ 150-220 Myr. The lifetime of a single a single star with mass $7M_{\odot}$ is $\sim$ 40 Myr. On average, the lifetime of the evolved star will be $\sim 4-5$ times longer than a single star of comparable mass. Hypervelocity binaries, as well as similar products captures around the MBH could thus form BSs. However, the latter binaries do not evolve in isolation, as they may be affected by encounters with other stars near the MBH (\citealt{Perets:09a}), or be affected by the MBH itself (Paper I). Further study of binary stellar evolution in the GC environment is required to predict the fate of such binaries.

\begin{table*}
\begin{tabular}{|r|r|r|r|r|r|}
\hline
outcome &$a_o/a_i=5$& $a_{o}/a_{i}=7.5$&$a_{o}/a_{i}=8.75$&$a_{o}/a_{i}=10$\\
\hline
HVS&0.62$\pm0.1$&0.56$\pm0.1$&0.35$\pm0.08$&0.36$\pm0.08$ \\
HVB&0.16$\pm0.0.05$&0.02$\pm0.02$&0.04$\pm0.03$&0$\pm0$\\
Collision&0.02$\pm0.02$&0.02$\pm0.02$&0.05$\pm0.03$&0.02$\pm0.02$ \\
Close Binary&0.2$\pm0.06$&0.27$\pm0.07$&0.27$\pm0.07$&0.33$\pm0.08$ \\
\hline
\end{tabular}
\caption{Probability of a HVS (first row), HVB (second row), collision (third row), and formation of a close binary (fourth row), from the disruption of a triple system of $4M_{\odot}$ stars with varying outer semimajor axis length. The semimajor axis of the inner binary is fixed at $0.05 AU$. This particular inner binary is very tight and thus has a relatively high probability of producing a HVB. The first column shows the probabilities of an outcome with an outer to inner semimajor axis ratio of 5. Note that this is the smallest stable outer semimajor axis length. The ratio of the semimajor axes in the second column is 1.5 times that of the first column. The ratios of the semimajor axes in the third and fourth columns are 1.75 and 2 times that of the first column respectively. All associated errors are Poisson. We can see that the production of HVSs drops as we get into larger ratios. However, at a certain distance the production of HVSs becomes solely due to the tight inner binary. The production of HVBs drops to $\sim$ 0 for larger outer semimajor axes. The collision rate remains constant within error, as does the production of close binary stars. We ran simulations with similar semimajor axis ratios but larger inner semimajor axis and results were similar to within the expected error.}
\label{tab_semimajor}
\end{table*}

\section{Conclusions} \label{Im}

In our study of triple disruptions by a MBH, we mapped the various possible outcomes, including the capture and ejection of single and binary stars (including HVSs and HVBs), collisions between two or all of the stars, and the perturbations of the orbital parameters of captured/ejected binaries. We also followed the long term stellar evolution of captured/ejected binaries which could lead to their merger.
  
Our N-body simulations indicate that if a triple system of inner binary separation $0.05$--$0.2$ AU with the third star at five times that distance approaches SgrA* to within $\la 10$ AU, the subsequent tidal disruption can lead to the production of a HVB. Such HVBs are always tight, and stellar evolution code shows that they evolve into massive BSs ($M_{final} \sim 7M_{\odot}$). It is possible that some of the bound HVSs are rejuvenated stars. However, it has not yet been confirmed that all these stars are true B-type stars and not horizontal branch stars. HVS HE 0437-5439, has been identified as a true young B2 III-IV halo star with mass $\sim 9M_{\odot}$ (\citealt{Bonanos:08}; \citealt{Przybilla:08}). Since the star likely originated from the GC (\citealt{Brown:10b}), it is quite possible that this HVS is a rejuvenated star, originally ejected as a binary from the GC (see Perets I).

In our simulations involving triples, the inner binary always has a mass twice that of the third companion. As expected, when a binary is ejected it typically has lower ejection velocity than compared with that of the companion (\citealt{Bromley:06}). The ejection of a HVB in our simulations is therefore less likely to occur than a HVS ejection, even though the inner binary and the third companion have a comparable chance of being ejected following the triple disruption. Nevertheless, the HVB ejection rate could extend to a non-negligible fraction of the HVSs (up to 1/4 of all HVSs in some of the cases we studied). A higher mass ratio between the third companion and the inner binary (as well as overall more massive triples, even with the same mass ratio) would lead to a higher HVB ejection rate.  The overall behavior is consistent with the assumptions used in previous studies of triple disruption (Perets I), and confirms their validity. One observed difference, however, is the possibility of a collision between the inner binary components, not accounted for in the previous studies of triple disruptions.  This interesting outcome reduces the HVB ejection rate, but only by a small fraction.

As mentioned above, in most cases a HVB is not formed upon the disruption of the triple system. However, even if a HVB is not formed, there is a fair chance that the resulting disruption will produce a tight binary around SgrA* (see Table \ref{tab_compare}). Barring tidal stripping or close encounters with massive systems, a tight binary orbiting SgrA* will evolve into a BS in the same manner as a HVB. It is possible that some of the massive S-stars are rejuvenated stars. It is also possible for a triple system to be disrupted so that two or all three stars collide. In some instances these collisions occur with $v < v_{esc}$ and thus the stars can coalesce and form a BS. If all three stars coalesce, this mechanism may explain some of the most massive stars we see around SgrA*.

In each scenario discussed in section \ref{FBS}, any star that remains orbiting SgrA* is always on a highly eccentric orbit with eccentricities ranging from $e = 0.9$ to almost unity. Such high eccentricities are inconsistent with the observed orbital parameters of the S-stars (e.g. \citealt{Gillessen:2009}). However,  resonant relaxation processes could change the eccentricity distribution over time, making the capture scenario a viable model for the origin of the GC S-stars (\citealt{Perets:07,Madigan:09,Perets:09a,Perets+:09a,Madigan:10,Perets:10}). We note that the initial eccentricity distribution of captured stars, following the binary/triple disruption in our simulations, suggests a lower eccentricity ($e=0.94$) than previously assumed for the initial conditions in simulations of the long term evolution of the S-stars (\citealt{Perets+:09a,Madigan:10}). Lower initial eccentricities are likely to provide better consistency between the observations and the simulation results.

Finally, this work covered only a small part of the possible phase space for triple disruption, but allowed us to map out their potential outcomes, and validate previous assumptions regarding triple disruption using detailed few-body simulations. Future extensions of this work may examine the disruption of triple systems of varying masses, inclinations, and eccentricities.

\section*{Acknowledgments}

We thank Avi Loeb and Gary Wegner for their useful suggestions. This work was supported in part by Dartmouth College and Harvard University funds.

\bibliographystyle{mn2e.bst}
\bibliography{v2Paper.bib}
\bsp

\end{document}